# Advanced Targeted Drug Delivery for Colon Cancer Using Pristine and Surface-modified Hydroxyapatite Nanoparticles: Synthesis, Characterization, and pH-Responsive Release

Alexander David McGuire Withrow[a], Sean M Blythe[b], Jack Thomas Burton[b], and Camryn Grace Evett[c]

[a]Department of Chemistry, Biochemistry and Physics, South Dakota State University, Brookings, SD, USA

[b]Department of Chemistry, Western Washington University, Bellingham, WA, USA

[c]Department of Medical Biology, University of South Dakota, Vermillion, SD, USA

Corresponding Author: Camryn Grace Evett; E-mail: camryn.evett@coyotes.usd.edu

**ABSTRACT**: Hydroxyapatite nanoparticles (HANs) have emerged as a promising candidate for targeted drug delivery in the treatment of colon cancer, a disease that remains one of the leading causes of cancer-related deaths globally. In this study, we synthesized and characterized HANs for their potential use as drug delivery vehicle for anticancer drugs specifically targeting colon cancer treatment. Through various in vitro assays, including drug loading efficiency, release kinetics, we demonstrated that HANs exhibit efficient drug loading capacity, and sustained release behavior, which could collectively contribute to enhanced therapeutic outcomes against colon cancer cells. The findings highlight the potential of HANs to serve as a versatile platform for the targeted drug delivery of chemotherapeutic agents, offering a novel approach that could improve treatment efficacy while minimizing side effects due to biocompatibility of the particles. Future research should focus on in vivo studies and clinical translation to further validate the potential of HANs in cancer therapy as drug carrier.

## 1. Introduction

Colon cancer is one of the most prevalent and deadly forms of cancer worldwide, representing a significant public health challenge. Despite advances in medical treatments, including surgery, chemotherapy, and radiation therapy, the prognosis for advanced-stage colon cancer remains poor. Conventional chemotherapy often lacks specificity, leading to systemic toxicity and adverse side effects, which limits its effectiveness. These challenges underscore the urgent need for more targeted and efficient drug delivery systems that can selectively attack cancer cells while sparing healthy tissues.

Nanotechnology has revolutionized the field of cancer therapy by offering innovative solutions to enhance drug delivery and therapeutic outcomes. Nanoparticles have gained considerable attention due to their ability to improve the solubility, colloidal stability, biocompatibility and bioavailability of anticancer drugs. Among the various types of nanoparticles explored, biocompatible and biodegradable materials are especially attractive for clinical applications. The ability to design nanoparticles with specific surface properties and functionalization allows for targeted delivery, reducing off-target effects and improving the therapeutic index.

Hydroxyapatite nanoparticles, which mimic the mineral component of bone, have emerged as a promising candidate for drug delivery applications due to their excellent biocompatibility, bioactivity, and ability to be functionalized for targeted therapy. HANs possess a porous structure that can be used to load and release drugs in a controlled manner, making them particularly suitable for delivering chemotherapeutic agents directly to cancer cells. Previous studies have demonstrated the potential of HANs in various biomedical applications, including bone regeneration and cancer therapy. However, their specific application in colon cancer treatment remains underexplored.

The objective of this study is to investigate the efficacy of hydroxyapatite nanoparticles as a targeted drug delivery system for colon cancer therapy. By characterizing the physicochemical properties of HANs, assessing their drug loading capacity, and evaluating their cytotoxic effects on colon cancer cells, we aim to demonstrate their potential as a novel therapeutic platform that could enhance the precision and effectiveness of cancer treatment while minimizing harmful side effects.

## 2. Materials and Methods

### 2.1 Synthesis of Hydroxyapatite Nanoparticles

Hydroxyapatite nanoparticles were synthesized using modified wet chemical precipitation method. In this process, calcium nitrate tetrahydrate ($Ca(NO_3)_2 \cdot 4H_2O$) and diammonium hydrogen phosphate (($NH_4)_2HPO_4$) were used as the calcium and phosphate precursors, respectively. Initially, a 0.4 M solution of calcium nitrate was prepared in deionized water and heated to 70°C under continuous stirring. Simultaneously, a 0.35 M solution of diammonium hydrogen phosphate was prepared. The phosphate solution was then slowly added dropwise to the calcium nitrate solution while maintaining constant stirring at 1000 rpm. The pH of the reaction mixture was adjusted to 10 using base solution ammonium hydroxide. The reaction was allowed to proceed for 2 h at 70°C, after which the precipitate formed was allowed to age overnight at room temperature. The resulting nanoparticles were collected by centrifugation at 10,000 rpm for 15 min, followed by multiple washes with deionized water and ethanol to remove any unreacted precursors and by-products. The particles were then dried at 80°C for 12 h in a vacuum oven. Finally, the dried HANs were calcined at 500°C for 3 h to improve crystallinity and remove any remaining organic residues.

### 2.2 PEGylation of hydroxyapatite nanoparticles

PEGylation of hydroxyapatite nanoparticles was done using a modified known procedure. To PEGylate hydroxyapatite nanoparticles, a 5mg/mL concentration solution of methoxy polyethylene glycol was prepared by dissolving in deionized water. Equimolar (1.0M) N-Hydroxysuccinimide (NHS) and 1-ethyl-3-(3-dimethylaminopropyl) carbodiimide (EDC) were added to this solution and stirred at room temperature for 30 min to activate the carboxyl groups of the mPEG. Hydroxyapatite nanoparticles were dispersed in deionized water by sonication for 15 min and then the mPEG solution was added dropwise to this suspension while stirring vigorously. After 24h of PEGylation reaction, PEGylated HANs were separated by centrifugation at 10,000 rpm for 10 min, followed by washing three times each with deionized water and ethanol to remove unreacted mPEG and by-products. Finally, the PEGylated nanoparticles were dried at 60°C under vacuum for 24 h to get the final product.

### 2.3 Characterization of Hydroxyapatite Nanoparticles

The synthesized HANs and PEG-HANs were characterized using a variety of techniques to determine their physicochemical properties. Transmission electron microscopy (TEM) was used to analyze the morphology and size of the HANs. Samples for TEM analysis was prepared by dispersing 1mg of the nanoparticles in 15mL of absolute ethanol and placed a drop on a carbon-coated copper grid, followed by drying on air. The surface charge of the HANs and PEG-HANs were measured using zeta potential analysis to ensure the PEGylation on the surface and assess the stability of particles in suspension. The nanoparticles (0.1 mg) were dispersed in 20 mL deionized water and analyzed the surface zeta potential at room temperature.

*2.4 Drug Loading and Release Studies*

The anticancer drug doxorubicin (DOX) was loaded onto the HANs and PEG-MSNs using a simple adsorption method. 10mg of particles was dispersed in 1mL of 0.5 mg/mL DOX solution by vortexing for 5 min and then stirred for 24 h at room temperature. Then drug-loaded particles were separated by centrifugation, washed two times with PBS (pH 7.4) to remove any unbound drug, the supernatants were kept for drug loading calculations. The drug loading efficiency was calculated by measuring the concentration of the free drug in the supernatant using UV-Vis spectroscopy at the characteristic absorption wavelength (480nm) of the drug doxorubicin in a microplate reader. In vitro drug release studies were conducted under different pH environments. Drug-loaded particles were suspended in phosphate-buffered saline (PBS) at pH 7.4 and placed in a shaking water bath at 37°C. At predetermined time intervals, samples were withdrawn, centrifuged, and the supernatant was analyzed for drug concentration using UV-Vis spectroscopy. The cumulative drug release was calculated and plotted as a function of time.

### 3. Results and Discussion

*3.1 Characterization of HANs and PEG-HANs*

The synthesized hydroxyapatite nanoparticles were characterized in detail to confirm their morphology and structural properties. Transmission Electron Microscopy was used to assess the size, shape, and surface structure of the nanoparticles. TEM images revealed that the HANs were predominantly spherical in shape with an average diameter ranging between 60-70 nm as shown in figure 1A. The nanoparticles demonstrated a uniform size distribution, indicating a controlled and consistent synthesis process. In addition to shape and size, high-magnification TEM images showed that the surface of the HANs was smooth and well-defined, with no significant agglomeration, which further confirmed the successful synthesis and stabilization of the nanoparticles.

Hydroxyapatite nanoparticles and PEGylated hydroxyapatite nanoparticles revealed surface zeta potential -20 mV and +30 mV respectively as shown in Figure 1B. The negative zeta potential of -20 mV for the unmodified HANs is primarily attributed to the presence of phosphate groups on the nanoparticle surface, which impart a negative charge under physiological pH conditions. This negative surface charge plays a crucial role in colloidal stability, as it induces electrostatic repulsion between particles, preventing aggregation and ensuring that the nanoparticles remain well-dispersed in aqueous suspension. The moderately high negative zeta potential suggests that the HANs have good colloidal stability, reducing the likelihood of agglomeration, which is essential for their effectiveness in biomedical applications. In contrast, the PEGylated hydroxyapatite nanoparticles exhibited a positive zeta potential of +30 mV, which can be attributed to the successful grafting of mPEG chains onto the nanoparticle surface. PEGylation alters the surface chemistry of the nanoparticles, likely by masking the negatively charged phosphate groups and introducing a net positive charge. The shift to a positive charge is indicative of successful PEG conjugation, which improves the hydrophilicity of the nanoparticles and enhances their colloidal stability. This is especially important for biomedical applications, as PEGylation increases circulation time in vivo and reduces the likelihood of opsonization by immune cells, promoting prolonged circulation and better therapeutic efficacy. Moreover, the positive zeta potential of the PEG-HANs facilitates potential interactions with negatively charged biomolecules or cell membranes, which could enhance cellular uptake and targeted drug delivery.

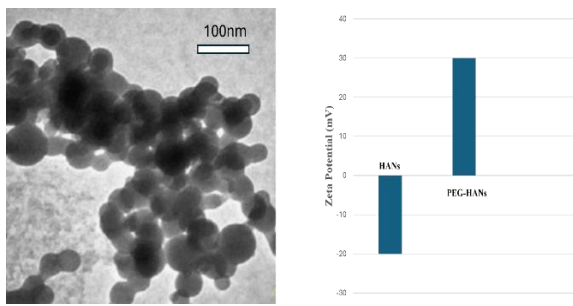

*Figure 1: Characterization of hydroxyapatite nanoparticles and PEGylated hydroxyapatite nanoparticles. TEM image showing morphology and size distribution (left), Zeta potential date indicating surface charge of pristine and PEGylated hydroxyapatite nanoparticles (right).*

*3.2 Drug Loading Efficiency and Release Profile*

The drug loading efficiency of hydroxyapatite nanoparticles (HANs) and PEGylated hydroxyapatite nanoparticles (PEG-HANs) was determined to be 78% and 89%, respectively, indicating the high capacity of both systems for drug adsorption. The porous structure and large surface area of the nanoparticles provide ample binding sites for the drug molecules. PEGylation further enhances the loading efficiency of HANs by improving their surface properties and hydrophilicity, facilitating better interaction between the drug and the nanoparticles. This higher efficiency in PEG-HANs can be attributed to the PEG chains, which create a favorable surface environment for drug interaction and retention.

In vitro drug release studies revealed a distinct pH-dependent release behavior in both systems, with significantly greater drug release occurring at pH 5.5 compared to pH 7.4. This indicates that both HANs and PEG-HANs respond effectively to acidic conditions, mimicking the tumor microenvironment, while exhibiting slower drug release at physiological pH (7.4). The biphasic release pattern observed in both cases includes an initial burst release followed by a sustained release phase. At pH 5.5, drug release was more rapid in both nanoparticle types due to the accelerated degradation of the nanoparticles under acidic conditions, which facilitates drug diffusion from the matrix.

Comparing the release profiles, PEG-HANs demonstrated a more controlled and prolonged drug release compared to HANs, both at pH 5.5 and pH 7.4. The PEGylation provides a stabilizing effect that slows down nanoparticle degradation and drug release, allowing for a more sustained and uniform release over time. At pH 5.5, the release from PEG-HANs was still substantial but more controlled, further indicating the modulatory role of PEG chains in nanoparticle stability and drug diffusion. This slower, sustained release is advantageous for maintaining therapeutic drug levels over a longer period, especially in tumor environments, reducing the need for frequent drug administration and potentially improving patient outcomes.

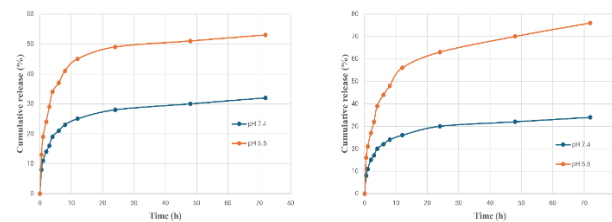

*Figure 2: In Vitro Drug Release Profiles from HANs and PEG-HANs. drug release profile for HANs at pH 5.5 & 7.4 (left), drug release profile for PEG-HANs at pH 5.5 & 7.4 (right).*

Overall, the pH-triggered drug release observed in HANs and PEG-HANs highlights their suitability as drug delivery systems, particularly for cancer therapy. The enhanced drug loading capacity, along with the controlled release behavior in response to acidic conditions, demonstrates the potential of these systems for targeted drug delivery, minimizing off-target effects and ensuring prolonged therapeutic efficacy. PEG-HANs, in particular, offer a promising platform for long-term drug release with better control over drug diffusion and nanoparticle stability.

## 4. Conclusion

This study demonstrates the successful synthesis and application of hydroxyapatite nanoparticles and PEGylated hydroxyapatite nanoparticles as efficient drug delivery systems for colon cancer therapy. Both pristine and surface modified nanoparticles exhibited excellent physicochemical properties, including uniform size distribution, high surface area, and colloidal stability. PEGylation further enhanced drug loading capacity (89% for PEG-HANs vs. 78% for HANs) and improved hydrophilicity. In vitro drug release studies showed pH-sensitive behavior, with significantly higher release in acidic conditions (pH 5.5), mimicking the tumor microenvironment, while PEG-HANs offered more controlled, sustained release. These findings highlight the potential of HANs and PEG-HANs as targeted, pH and enzyme-responsive drug delivery systems with improved therapeutic efficacy, biocompatibility and reduced toxicity, paving the way for future preclinical and clinical investigations in colon cancer treatment.